# Deep Learning Segmentation and Classification of Red Blood Cells Using a Large Multi-Scanner Dataset

Mohamed Elmanna [a], Ahmed Elsafty [b], Yomna Ahmed [b], Muhammad Rushdi [a, c*], and Ahmed Morsy [a]

[a] Department of Biomedical Engineering and Systems, Faculty of Engineering, Cairo University, Cairo University St, Giza 12613, Egypt
[b] PathOlOgics, LLC for processed medical data and tele-pathology services, Egypt
[c] School of Information Technology, New Giza University, Giza 12256, Egypt

**Abstract**

Digital pathology has recently been revolutionized by advancements in artificial intelligence, deep learning, and high-performance computing. With its advanced tools, digital pathology can help improve and speed up the diagnostic process, reduce human errors, and streamline the reporting step. In this paper, we report a new large red blood cell (RBC) image dataset and propose a two-stage deep learning framework for RBC image segmentation and classification. The dataset is a highly diverse dataset of more than 100K RBCs containing eight different classes. The dataset, which is considerably larger than any publicly available hematopathology dataset, was labeled independently by two hematopathologists who also manually created masks for RBC cell segmentation. Subsequently, in the proposed framework, first, a U-Net model was trained to achieve automatic RBC image segmentation. Second, an EfficientNetB0 model was trained to classify RBC images into one of the eight classes using a transfer learning approach with a 5×2 cross-validation scheme. An IoU of 98.03% and an average classification accuracy of 96.5% were attained on the test set. Moreover, we have performed experimental comparisons against several prominent CNN models. These comparisons show the superiority of the proposed model with a good balance between performance and computational cost.

## 1. Introduction

The complete blood count (CBC) test is one of the most valuable medical tests that provides important information about the cellular blood components. The value of the CBC test, however, is limited by the fact that this test only offers minimal information about blood cell abnormalities. As a result, if abnormal qualitative or quantitative signals are found in a CBC test, a peripheral blood smear test may be necessary. For example, proper classification of abnormal red blood cells (RBCs) could be critical because cell abnormalities are closely linked to various disease changes [1]. Blood smear analysis has traditionally been performed through human inspection, which is time-consuming, requires well-trained professionals, and is vulnerable to subjectivity and intra-observer variability [2].

Computer-aided techniques have been proposed to reduce the limitations of manual analysis [3]. Machine learning methods, especially deep learning (DL) ones, have outperformed conventional computer-aided diagnostic techniques and sped up progress in biomedical image analysis [4]. In particular, deep neural networks are now among the most extensively utilized machine learning approaches for medical image analysis tasks, such as image detection [5]-[7], enhancement [8], [9], segmentation [10]-[12], and classification [13]-[17].

As a result, digital pathology (DP) has emerged as a promising field based on advancements in artificial intelligence, deep learning, high-performance computing, and large-scale data analytics. In this paper, we introduce a large-scale medical image dataset for red blood cell (RBC) segmentation and classification. Moreover, we propose an advanced two-stage DL framework for RBC segmentation and classification. This framework was validated and compared against several methods based on other state-of-the-art architectures, such as ResNet, ConvNeXt, and MobileNet.

## 2. Related Work

Automated analysis of red blood cells has been extensively investigated in the literature in recent years. Common tasks include cell localization and counting [18]-[20], cell segmentation [21]-[23], and cell classification [24]-[27]. Nevertheless, one of the major challenges of earlier work on RBC classification is the limited availability of data [24], [25], [27], [28]. Until recently, ErythrocytesIDB [29] was the only publicly available dataset for RBC morphology classification. This dataset contains only 629 cell images for 3 RBC classes (round, oval, and others). Tyas et al. [25] used a multi-layer perceptron to classify nine RBC types present in thalassemia cases. A dataset of 7,108 RBC samples was used to train RBC classifiers based on combinations of morphological, texture, and color features.

Naruenatthanaset et al. [24] proposed a method for RBC segmentation and classification into 12 RBC classes. Firstly, a color normalization step was employed to reduce color variance in the RBC images. Subsequently, an EfficientNetB1 classifier was trained and tested on a dataset of 20,875 RBC samples, giving a classification accuracy of 92.1%. Durant et al. [28] collected a dataset of 3,737 images for RBC classification with 10 classes. A DenseNet architecture was then trained and tested, and a classification accuracy of 90.6% was attained. For sickle-cell anemia diagnosis, Alzubaidi et al. [27] proposed deep learning models to classify RBCs into three classes. These models were supported by schemes of same-domain transfer learning and data augmentation. Model training was carried out on the ErythrocytesIDB dataset [29], and a classification accuracy of 99.54% was reported. Furthermore, several studies focused on RBC classification for disease-specific tasks, such as malaria parasite detection [30], malaria life-cycle classification [31], and thalassemia detection [25], [32].

---
* Corresponding author.
*E-mail address:* mrushdi@eng1.cu.edu.eg (Muhammad Rushdi)

## 3. Existing Datasets of Blood Cell Images

Building a good machine learning model requires a high-quality dataset of a suitable size. We review here the characteristics and limitations of existing publicly available blood smear image datasets for cell detection, segmentation, and classification. Table 1 summarizes the key characteristics of these datasets. Although some of these datasets are for white blood cells (WBCs) only, they are still included in the table for completeness.

### 3.1. Existing publicly available blood cell image datasets

**Blood Cell Count and Detection Dataset (BCCD)** [33]: This is a small-sized dataset of 364 images used primarily for cell detection. The dataset includes 4,888 bounding-box annotations for three cell types (RBCs, WBCs, and platelets) where the majority of cells is of the RBC type. Each image is of the JPEG format with a size of $640 \times 480$ pixels.

**Raabin-WBC** [34]: This WBC image dataset was released in 2021 for WBC classification. Each cell image was cropped and assigned by two experts into one of five classes, namely, mature neutrophils, lymphocytes (small and large), eosinophils, monocytes, and basophils. The dataset contains about 40,000 WBC images acquired from two scanners. Also, for cytoplasm and nucleus segmentation, 1,145 ground-truth masks were created by human experts.

**Microscopic peripheral blood cell image dataset** [35]: This dataset contains 17,092 normal WBC images for WBC classification. The data samples were annotated by qualified clinical pathologists into neutrophils, eosinophils, basophils, lymphocytes, monocytes, immature granulocytes (promyelocytes, myelocytes, and metamyelocytes), erythroblasts, and platelets or thrombocytes. The images are in the JPEG format with dimensions of $360 \times 363$ pixels.

**Leukocyte Images for Segmentation and Classification (LISC)** [36]: In this dataset, 100 microscope slides were prepared for the peripheral blood of 8 healthy human subjects. The microscope slides were smeared and stained using the Gismo-Right method, and then 400 image samples were captured from the stained peripheral blood using an achromatic lens and a light microscope. The images were saved in the BMP format with a size of $720 \times 576$ pixels. Each image was labeled by a human expert into one of five normal leukocyte types: basophil, eosinophil, lymphocyte, monocyte, and neutrophil. Also, 250 ground-truth nucleus and cytoplasm masks were created by the human expert for segmentation tasks.

**RBCdataset** [25], [37]: This dataset was acquired from four thalassemia blood smears and a healthy blood smear. The images were acquired by an Optilab Advance Plus camera with an Olympus CX21 microscope. The dataset contains 7,108 grayscale cell images of nine RBC types (elliptocyte, pencil cell, teardrop, acanthocyte, stomatocyte, target cell, spherocyte, hypochromic, and normal cell). The images are in a PNG format, and the image sizes vary according to the cell size.

**Chula RBC-12-Dataset** [24]: To the best of our knowledge, this is the largest public RBC image dataset for cell detection and classification tasks. It contains 706 smear images with over 20K labeled RBC cells from 12 RBC types (normal cell, macrocyte, microcyte, spherocyte, target cell, stomatocyte, ovalocyte, teardrop, burr cell, schistocyte, hypochromia, and others). Each smear image has a size of $640 \times 480$ pixels and ground-truth information of cell locations ($x$ and $y$ coordinates of cell centers) and cell types.

**ErythrocytesIDB** [29]: This dataset is used to perform cell segmentation and classification. The dataset contains peripheral blood smear images taken from patients with sickle cell disease. The dataset has three subsets for different tasks. The first subset has 196 smear images and 629 RBC images of three classes (cellular, elongated, and other). The second and the third subsets contain segmentation masks for 50 and 30 smear images, respectively.

### 3.2. Limitations of existing datasets

Each of the aforementioned datasets comes with some limitations. The BCCD [33] is a small dataset with 4888 samples unevenly distributed among RBCs (4155), WBCs (372), and platelets (361). The Raabin-WBC dataset [34] is the most-recent largest dataset for WBC classification, but it contains a small number of cytoplasm and nucleus segmentation masks. The microscopic peripheral blood cell image dataset [35] was collected from only one scanner. The ErythrocytesIDB [29] dataset is small and also comes from one source. The LISC dataset [36] is small, has a limited number of segmentation masks, and is acquired from only one imaging source. The RBCdataset [25], [37] is also a small recent dataset with only 7,108 grayscale RBC images acquired using one camera. Although the Chula RBC-12-Dataset [24] has several variations of RBC types with relatively large numbers of images, no segmentation masks are available, and all images were acquired from only one microscope. All these limitations emphasize the need for building a large-scale RBC image dataset with more realistic variations, annotation labels for RBC types, and ground-truth masks for cell segmentation. Such a dataset can have a significant impact on the research in RBC segmentation and classification. We introduce in this work a large dataset that meets these criteria.

## 4. Methodology

Our goal in this paper is to build a two-stage deep learning framework for RBC image segmentation and classification using a large-scale RBC dataset. Fig. 1 illustrates the detailed steps of the proposed framework. First, a large and diverse dataset of RBC images with eight classes was collected, processed, segmented, and independently labelled by two hematologists with a combined medical experience of over 20 years as well as seven years of data science practice. Then, a U-Net deep-learning model was trained on a subset of the collected dataset to perform RBC image segmentation (See Fig. 1a). After completing the segmentation stage, an EfficientNetB0 deep architecture was trained on another subset of the dataset for RBC image classification (See Fig. 1b). In the following, we provide more details on data collection (Section 4.1), RBC image segmentation (Section 4.2), and RBC image classification (Section 4.3).

**Table 1**
List of the publicly available datasets for cell detection, segmentation, and classification.

| Dataset | Tasks | # of images | Cell types | Sub-types | Image format | Width × Height | Source | Released |
|---|---|---|---|---|---|---|---|---|
| **Blood Cell Count and Detection Dataset (BCCD)** [33] | Cell detection | 364 images | RBCs, WBCs, Platelets | NA | jpg | 640×480 | NA | 2018 |
| **Raabin-WBC** [34] | Cell classification / Cytoplasm and nucleus segmentation | 40,000 cell images / 1,145 cell images | WBCs | 5 | jpg, bmp | variable | Two scanners | 2021 |
| **Microscopic peripheral blood cell image dataset** [35] | Cell classification | 17,092 cell images | WBCs | 8 | jpg | 360×363 | One scanner | 2020 |
| **LISC: Leukocyte Images for Segmentation and Classification** [36] | Cell classification / nucleus and cytoplasm segmentation | 400 smear images / 250 smear images | WBCs | 5 | bmp | 720×576 | One camera | 2011 |
| **RBCdataset** [25, 37] | Cell classification | 7,198 cell images | RBCs | 9 | png (grayscale) | variable | One camera | 2022 |
| **Chula RBC-12-Dataset** [24] | Cell detection & classification | 738 smear images, 20,875 cell images | RBCs | 12 | jpg | 640×480 | One camera | 2021 |
| **ErythrocytesIDB** [29] | Cell segmentation / Cell classification / Cell segmentation | 196 smear images / 629 cell images / 50, 30 smear images | RBCs | 3 | jpg | variable | One camera | 2017 |
| **100K-RBC-PathOlOgics dataset (introduced in this paper)** | **Cell classification** | **100,873 cell images** | RBCs | 8 | jpg | 80×80 | **Four scanners** | 2023 |
| **100K-RBC-Mask-PathOlOgics dataset (introduced in this paper)** | **Cell segmentation** | **100,118 cell images** | | | | | | |

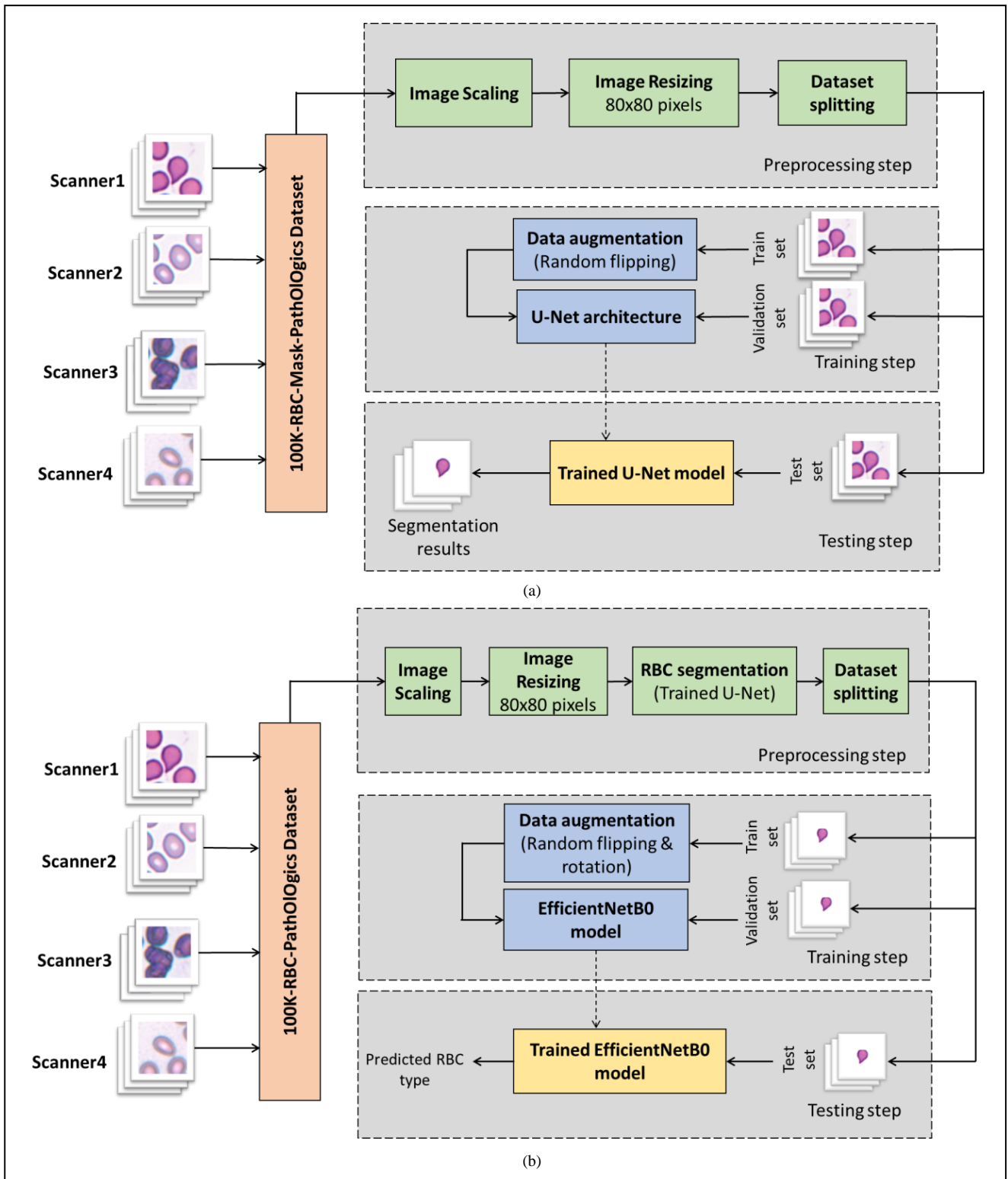

**Fig. 1.** The general framework for the proposed red blood cell segmentation and classification systems, (a) The proposed RBC segmentation system using a U-Net architecture, (b) The proposed RBC classification system using an EfficientNetB0 architecture.

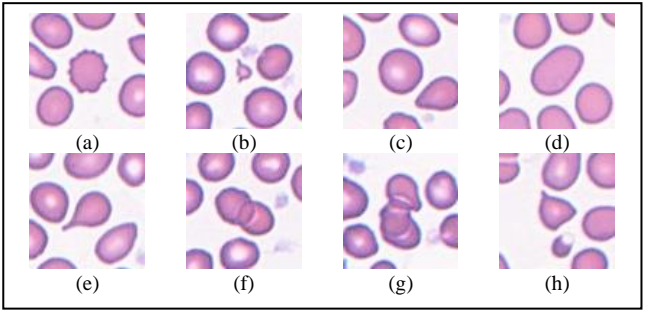

**Fig. 2.** Samples of eight RBC types. (a) Burr cells, (b) Fragmented RBCs, (c) Normal RBCs, (d) Ovalocytes, (e) Teardrops, (f) Two overlapped RBCs, (g) Three overlapped RBCs, and (h) Other RBC types.

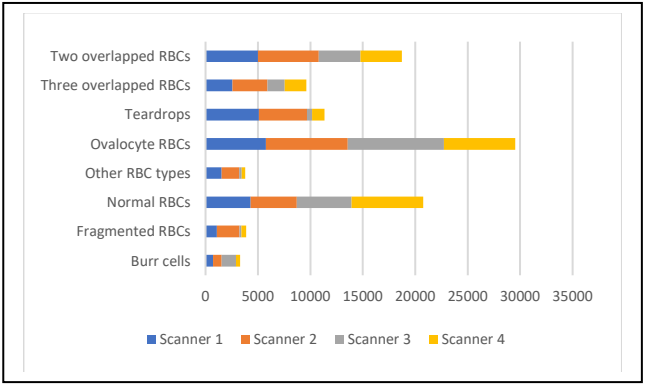

**Fig. 3.** Distribution of the 100K cells among eight RBC classes and four scanner types.

### 4.1. 100K-RBC-PathOlOgics: Multi-scanner RBC dataset

The two hematologists curated large datasets of RBC images, each with its corresponding ground-truth binary semantic segmentation mask. They selected 25 manually prepared and stained peripheral-blood and bone-marrow smears from patients suspected to have primary myelofibrosis (PMF) of the bone marrow. For generating whole-slide images (WSIs), four digital pathology scanners were utilized. Each scanner was equipped with a different light microscope using 40X magnification power, control units, integrated cameras, and software for patch stitching. The 25 slides were divided into 4 subsets and each scanner was used to scan a different subset. Table 2 illustrates the distribution of the 25 slides among the four scanners. The physical dimensions of each slide was about 1×2 inches. Each slide was subdivided into around 2,000 non-overlapping patches with a size of 539×1076 pixels. Each patch had more than 200 blood cells. To create masks, the hematologists developed a custom semi-automatic segmentation scheme as follows. First, the cellular borders are manually outlined using a digital pen. Then, the identified cell within a cropped image is automatically centered and the mask is precisely adjusted to match the central position of the cell.

The first dataset, named *100K-RBC-Mask-PathOlOgics*, comprised 100,118 cropped RBC images, each paired with its corresponding mask. This dataset was used for developing automated RBC cell segmentation algorithms. The second dataset, named *100K-RBC-PathOlOgics*, included 100,873 segmented RBC images and was utilized for training and testing RBC classifiers.

The labeling criteria were designed to focus on clinically significant RBC types, where visual examination is considered exclusive and unassisted by current technological solutions. After each hematologist independently labeled the complete dataset, a final discussion was conducted to address and resolve any discrepancies in the labeling process. Each cell was categorized into one of eight classes, namely, normal/rounded RBCs, ovalocytes (oval or egg-shaped), burr cells (crenated), fragmented RBCs, teardrop-shaped RBCs, two overlapped RBCs, three overlapped RBCs, and other RBCs that contain artificial/false teardrops. The presence of ovalocytes with more than 5% of the total RBC count is associated with almost all types of anemia. Additionally, teardrops and fragmented RBCs are linked to serious medical conditions and their presence should never be underestimated, as they could be fatal. Moreover, burr cells emerge due to medical causes or a poor staining process. Lastly, the utilization of manually prepared smears, which is prevalent in medical labs worldwide, may result in the unsuitability of whole-slide image (WSI) areas for RBC examination and counting. In fact, the selection of optimal counting areas depends on evaluating the ratio between the counts of the individual cells and the overlapped cells. Regions with a lower occurrence of overlapped cells are considered more appropriate for accurate examination and counting purposes.

This dataset has several key advantages. First, because of the sample diversity within the dataset, a well-trained classifier can effectively operate on manually prepared and stained smears without mandating prior standardization of staining or smearing. Also, such a classifier can serve as a sensitive screening tool for anemia, based on the percentage of the ovalocytes among all RBCs, while also exhibiting high specificity for identifying teardrops and fragmented RBCs. At the same time, this classifier can use burr cells percentages to detect improper manual staining with high sensitivity. Furthermore, the classifier can identify the most suitable areas to initiate cell counting, relying on the ratio between the overlapped and individual cells.

Samples of the RBC images from the eight classes are shown in Fig. 2. Fig. 3 shows the distribution of the collected images among the eight RBC classes and Fig. 4 presents samples of RBC images with their corresponding segmentation masks. Notably, each of the selected 25 smears contained samples from all targeted RBC classes, with variations in class proportions. As a result, images representing every RBC class were successfully collected from each WSI. Fig. 5 illustrates the appearance and color variations among images collected by the four scanners.

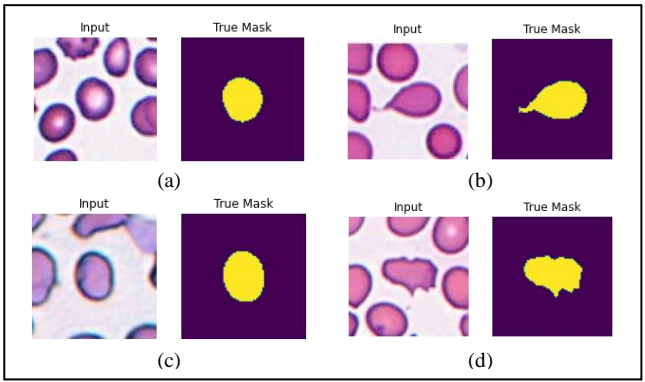

**Fig. 4.** Samples of RBCs with their corresponding ground-truth masks. (a) Normal RBC, (b) Teardrop, (c) Ovalocyte, and (d) Burr cell.

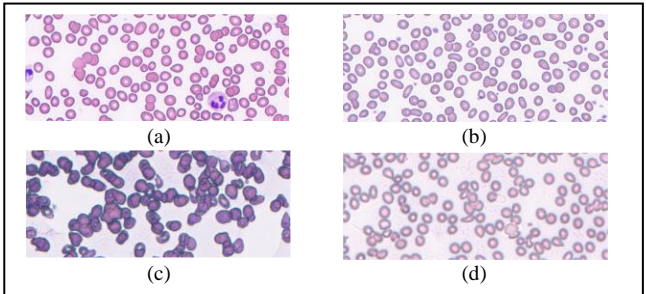

**Fig. 5.** RBC image patches collected from four different scanners: (a) Images from Scanner 1, (b) Scanner 2, (c) Scanner 3, and (d) Scanner 4. These images are cropped versions from the whole patch images to clearly visualize the RBCs.

**Table 2**
Distribution of the 25 WSI slides among the four scanners.

| Scanners no. | WSI slide no. | | | | | | |
|---|---|---|---|---|---|---|---|
| **1** | 1 | 5 | 6 | 8 | 25 | | |
| **2** | 2 | 3 | 4 | 7 | 9 | 11 | |
| **3** | 14 | 15 | 19 | 20 | 22 | 23 | 24 |
| **4** | 10 | 12 | 13 | 16 | 17 | 18 | 21 |

### 4.2. Semantic segmentation of RBCs

For the RBC image segmentation stage, we used the U-Net architecture which is considered one of the popular methods for medical image segmentation [38]. The U-Net architecture consists of contracting and expansive paths. The contracting path consists of two 3×3 convolutions that are applied repeatedly, where each convolution is followed by a rectified linear unit (ReLU), batch normalization, and a 2×2 max-pooling step for downsampling. The number of feature channels is doubled for each downsampling step. Each step in the expansive path involves upsampling of the feature map followed by a 2×2 convolution that reduces the number of feature channels by half. This convolution is followed by a concatenation with the corresponding feature map from the contracting path, two 3×3 convolutions, and ReLU activation functions [38]. The 100K-RBC-Mask-PathOlOgics dataset was used for U-Net training with corresponding ground-truth masks (Section 4.1). This dataset was normalized to the range (0-1), resized to 80×80 pixels, and split into three independent training, validation, and testing subsets with RBC images from 15 slides (60%), 5 slides (20%), and 5 slides (20%), respectively. All scanner types and RBC types were represented in the three sets. In the training phase, training images were passed to a data augmentation block to randomly flip all images vertically and horizontally on-the-fly, without actually adding more images to the training data.

### 4.3. Classification of RBCs

For RBC image classification, we used transfer learning with the pre-trained EfficientNetB0 network [39], which was already trained on the ImageNet dataset. EfficientNet is a family of CNN architectures developed using uniform scaling of the network depth, width, and resolution with a set of scaling coefficients. In particular, EfficientNetB0, the core variant of the EfficientNet family, was built using a multi-objective neural architecture search technique that optimizes accuracy and FLOP count [39]. This led to model with strong ability of reaching a suitable balance between accuracy and computational efficiency.

The 100K-RBC-PathOlOgics dataset was used for classifier training with eight RBC classes (Fig. 2). RBC images from five slides (20% of the slides) were used as the test subset, while the remaining slide images were split using 5×2 cross-validation into training and validation subsets. This division scheme ensures independence between the three subsets as each subset includes samples from distinct slides (or patients). In the training phase, the training image count was increased through on-the-fly data augmentation operations (vertical and horizontal random flipping, as well as random image rotations with angles between ±36 degrees). Based on these operations, random perturbations were applied to all training images during each epoch and the perturbed images were added to the original training data in order to prevent model overfitting. For loss computations, we used the sparse cross-entropy loss in the proposed model. Also, we tried a class-weighted loss to deal with the class imbalance problem. For the class-weighted loss, the weight of each class was set based on the number of samples in that class:

$$Class\ weight = \frac{Total\ number\ of\ samples}{No.\ of\ class\ samples \times No.\ of\ classes} \quad (1)$$

### 4.4. Implementation

All deep learning models were implemented using the Keras and TensorFlow libraries. The Adam optimizer was used for model optimization. The learning rate (LR) was initialized to 0.0004 for both the segmentation and classification models. The LR was reduced by a factor of 10 if the loss did not improve by at least 0.0001 for four epochs. The training was stopped if no loss reduction occurred for 10 epochs. The model with the minimum validation loss was selected as the best model. A batch size of 32 was used, and all input images were normalized between 0 and 1.

## 4.5. Evaluation measures

To evaluate the segmentation and classification models, we used the following metrics: sensitivity (recall), specificity, precision (positive predictive value (PPV)), negative predictive value (NPV), F1-score, accuracy, the false positive rate, and the intersection over union (IoU).

### 4.5.1. Segmentation performance measures

In a segmentation task, the true-positive ($TP$) count is the number of pixels correctly predicted as foreground pixels, while the false-positive ($FP$) count is the number of pixels actually belonging to the background but misclassified as foreground pixels. Similarly, the false-negative ($FN$) count is the number of foreground pixels misclassified as background pixels. The true-negative ($TN$) count is the number of pixels that belong to the background and were classified as such.

Sensitivity ($SN$) or recall is the percentage of foreground pixels actually classified as such.

$$SN = \frac{TP}{TP+FN} \quad (2)$$

Specificity ($SP$) the percentage of background pixels actually classified as such.

$$SP = \frac{TN}{TN+FP} \quad (3)$$

Accuracy ($ACC$) is the percentage of the correctly classified pixels (either foreground or background pixels) among all pixels.

$$ACC = \frac{TP+TN}{TP+TN+FP+FN} \quad (4)$$

Precision or the positive predictive value ($PPV$) is the number of correctly predicted foreground pixels to the total number of pixels predicted as foreground.

$$PPV = \frac{TP}{TP+FP} \quad (5)$$

The intersection over union ($IoU$), a measure of segmentation performance, is the ratio of overlapped between the predicted segmentation output and the ground-truth segmentation.

$$IoU = \frac{TP}{TP+FP+FN} \quad (6)$$

The F1-score ($F_1$) is the harmonic mean of the precision and recall. This measure is also used in classification tasks.

$$F_1 = 2 \times \frac{PPV \times SN}{PPV+SN} = \frac{2 \times TP}{2 \times TP+FP+FN} \quad (7)$$

The false positive rate ($FPR$) measures the ratio between the number of misclassified background pixels and the total number of background pixels.

$$FPR = 1 - SP = \frac{FP}{FP+TN} \quad (8)$$

### 4.5.2. Classification performance measures

In a binary classification task for a cell class X (against cells from all other classes), the true-positive ($TP$) count is the number of class-X samples correctly predicted as such, while the false-positive ($FP$) count is the number of samples actually belonging to other classes but misclassified to be in class X. Similarly, the false-negative ($FN$) count is the number of class-X samples that were misclassified into other classes. The true-negative ($TN$) samples are the ones that don't belong to class X and were classified as such.

Sensitivity ($SN$) or recall is the percentage of the class-X samples correctly classified as such (See Eq. (2)).

Specificity ($SP$) is the percentage of samples of any class other than X, correctly classified as non-class-X samples (See Eq. (3)).

Accuracy ($ACC$) is the proportion of the correctly classified samples (from all cell classes) among all samples (See Eq. (4)).

Precision or the positive predictive value ($PPV$) is the ratio of the number of samples correctly predicted as positive to the total number of samples predicted as positive (See Eq. (5)).

The negative predictive value ($NPV$) is the ratio of the number of samples correctly predicted as negative to the total number of samples predicted as negative.

$$NPV = \frac{TN}{TN+FN} \quad (9)$$

## 5. Experimental Results

### 5.1. Evaluation of RBC segmentation

In this section, we present the performance evaluation results for the deep-learning-based RBC segmentation model. Table 3 presents the evaluation measures for the U-Net model on the test set (18,690 RBC images extracted from five WSIs). the trained model achieved promising results with F1-score and IoU of 99.01% and 98.03%, respectively.

**Table 3**
Performance Evaluation measures for the U-Net RBC segmentation model on the test set (18,690 RBC images extracted from five WSIs).

| Measure | Value (%) |
|---|---|
| Sensitivity ($SN$) | 98.75 |
| Specificity ($SP$) | 99.89 |
| Precision ($PPV$) | 99.27 |
| F1-score ($F_1$) | 99.01 |
| Accuracy ($ACC$) | 99.73 |
| False positive rate ($FPR$) | 0.11 |
| Intersection over union ($IoU$) | 98.03 |

**Table 4**
Average performance evaluation results for the proposed RBC classification model with unweighted and class-weighted loss functions using a 5×2 cross-validation scheme.

| Model | Metrics | RBC types | | | | | | | | Average |
|---|---|---|---|---|---|---|---|---|---|---|
| | | OTH | BUR | FRA | NOR | OVA | TEA | THR | TWO | |
| **Proposed method + unweighted loss** | SN (%) | 76.12 | 95.34 | 99.08 | 97.78 | 97.21 | 94.61 | 94.52 | 97.76 | **94.05** |
| | SP (%) | 99.52 | 99.79 | 99.92 | 99.27 | 98.72 | 99.56 | 99.66 | 99.25 | **99.46** |
| | F1 (%) | 78.72 | 93.77 | 98.67 | 97.80 | 97.08 | 94.39 | 95.67 | 97.26 | **94.17** |
| | PPV (%) | 81.57 | 92.35 | 98.27 | 97.81 | 96.95 | 94.17 | 96.87 | 96.76 | **94.34** |
| | NPV (%) | 99.34 | 99.88 | 99.96 | 99.26 | 98.83 | 99.60 | 99.39 | 99.49 | **99.47** |
| **Proposed method + class-weighted loss** | SN (%) | 86.20 | 96.68 | 99.17 | 98.20 | 94.65 | 89.98 | 94.85 | 96.94 | **94.58** |
| | SP (%) | 98.77 | 99.60 | 99.93 | 98.93 | 99.13 | 99.73 | 99.53 | 99.31 | **99.36** |
| | F1 (%) | 75.01 | 90.98 | 98.81 | 97.51 | 96.22 | 92.97 | 95.30 | 96.95 | **92.97** |
| | PPV (%) | 66.67 | 85.95 | 98.45 | 96.84 | 97.86 | 96.21 | 95.82 | 96.98 | **91.85** |
| | NPV (%) | 99.61 | 99.92 | 99.96 | 99.40 | 97.78 | 99.25 | 99.43 | 99.30 | **99.33** |

Note. SN = Sensitivity; SP = Specificity; F1 = F1-score; PPV = Precision; NPV = Negative predictive value; OTH = Other RBC types; BUR = Burr cells; FRA = Fragmented RBCs; NOR = Normal RBCs; OVA = Ovalocytes; TEA = Teardrops; THR = Three overlapped RBCs; TWO = Two overlapped RBCs

### 5.2. Evaluation of RBC classification performance

In this part, we present the RBC classification results for the proposed model. An EfficientNetB0 architecture is trained on the classification dataset with unweighted and class-weighted losses using a 5×2 cross-validation scheme.

Table 4 shows the average performance measures on the test set for the proposed model with the two loss types. The results show the ability of the proposed model to achieve high average sensitivity and average F1-score of 94.05% and 94.17%, respectively. Moreover, the proposed model with a class-weighted loss leads to a higher average sensitivity of 94.58% and a lower average F1-score of 92.97%.

The results reveal that the proposed model with the unweighted loss can achieve good performance for fragmented RBCs, normal RBCs, ovalocytes, and two overlapped cells with sensitivity values of 99.08%, 97.78%, 97.21%, and 97.76%, respectively. In addition, three classes (burr cells, teardrops and three overlapped cells) have high sensitivity values of more than 94%. However, inferior performance is observed for the other RBC types, with sensitivity and precision values of 76.12% and 81.57%, respectively.

The proposed model with class-weighted loss achieves higher sensitivity for the minority classes (burr cells and other RBC types) with sensitivity values of 96.68% and 86.20%, respectively. However, that model shows a clear reduction in precision for burr cells and other RBC types as well as a reduction in sensitivity for teardrops and ovalocyte RBCs compared to the proposed model with the unweighted loss.

For the proposed system with the unweighted loss, the confusion matrix (averaged over the 5×2 cross-validation folds) is shown in Fig. 6a. Obviously, the highest confusion is between the normal RBCs and the ovalocytes. Also, there is a clear confusion between two and three overlapped cells. Another significant confusion is between teardrops and other RBCs.

The confusion matrix for the proposed model with the class-weighted loss is presented in Fig. 6b. There is a clear improvement in the sensitivity by 10.1% for the other RBC types minority class. The sensitivity still dropped by 4.6% and 2.6% for the teardrop and ovalocyte classes, respectively. For the other classes, smaller sensitivity improvements can be observed.

In Fig. 7, we present average ROC curves of the proposed classification model with the unweighted loss for the eight RBC classes using the full range for the false-positive rate on the x-axis and the true-positive rate on the y-axis (Fig. 7a). In Fig. 7b, we present a magnified version of the ROC curves in the range of [0.8, 1] for the true-positive rate and the range of [0, 0.2] for the false-positive rate. We can see that the largest AUC value (AUC = 0.999) is achieved by the fragmented RBCs (with only 7 misclassified samples out of 818 ones) while the lowest AUC value (AUC = 0.9876) is associated with the other RBC types (with 117 misclassified samples out of 488 ones).

Moreover, we compare the performance of our RBC classification model against eight state-of-the-art CNN models: ResNet50 [40], ResNet50V2 [41], Xception [42], DenseNet121 [43], MobileNetV3Small [44], MobileNetV3Large [44], ConvNeXtTiny [45], and EfficientNetV2B0 [46]. The experimental settings for our model and all these models are the same except for the employed classification module. In particular, the hyperparameters, the training set, and the test set used for building all models are the same to ensure a fair comparison. We train and test all models using the Google Colaboratory framework on a Tesla T4 GPU (with a 16-GB RAM) and an Intel(R) Xeon(R) CPU (with a 2.3-GHz microprocessor and a 12.7-GB RAM).

Table 5 presents the test accuracy, sensitivity, precision, F1-score, the training time, the testing time on a GPU, the number of trained model parameters, and the storage size. Also, Table 6 presents the results of two-sample t-tests for the statistical significance of the performance differences between the proposed model and each of the eight SOTA CNN models.

All models reach high accuracy levels in the range of 93.4% - 96.7% with F1-scores in the range of 89.7% - 94.5%. The MobileNetV3 models were designed to be lightweight models for mobile vision applications, and this is why these models understandably outperform ours in terms of computational cost measures. However, our proposed model is significantly better than the MobileNetV3Large and MobileNetV3Small models in terms of the four performance measures. ConvNeXtTiny, the smallest architecture in the ConvNext model family, outperforms all other models in terms of the four performance measures. Still, our model achieves comparable results with only a 0.3% difference in the average F1-score. Also, ConvNeXtTiny is the worst model for all computational cost measures.

Although the Xception model shows better performance compared to our model, our model narrowly matches this performance with a small difference of the F1-score of only 0.2%. Also, our model is significantly better than the Xception model in terms of all computational cost measures with percentage drops in training time, GPU-based testing time, number of trained parameters, and storage size, of 16.5%, 39.8%, 80.6%, and 80.3%, respectively.

In comparison with EfficientnetV2B0, our proposed model shows a significant improvement in the testing time with insignificant differences of the four classification measures between the two models.

In Fig. 8, we present a visual comparison of the performance and computational cost measures of the proposed model against the eight CNN models. Obviously, our model reaches a good balance between the computational cost and the classification performance.

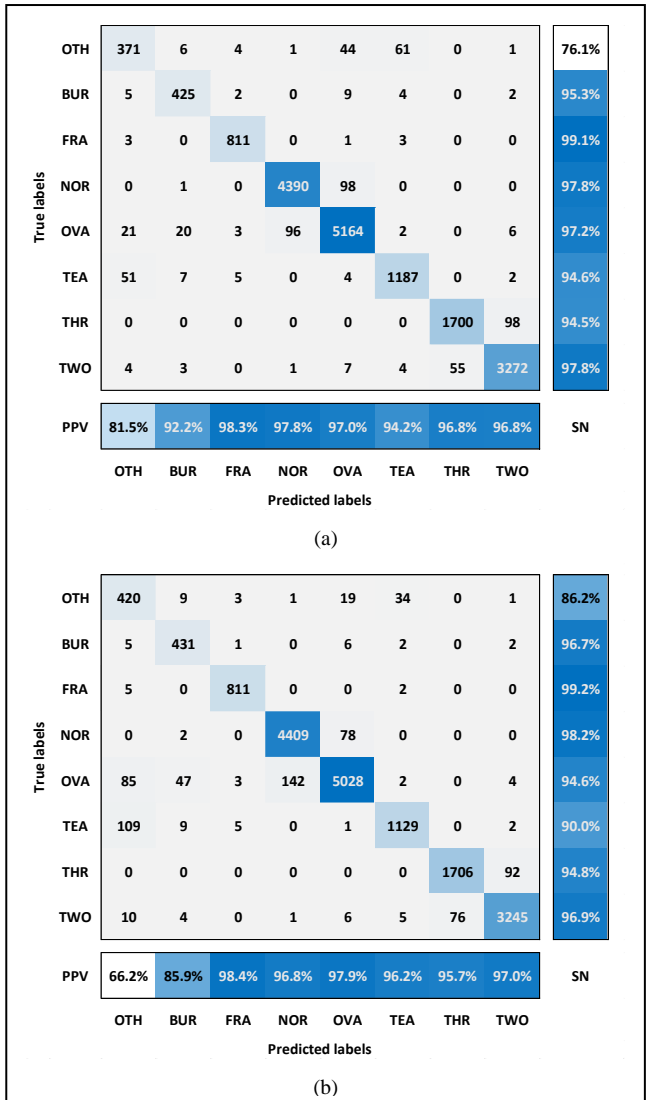

**Fig. 6.** Confusion matrices for the proposed RBC classification model. (a) the proposed model + unweighted loss, (b) the proposed model + class-weighted loss. The reported percentages are averages of the ones obtained from the 5×2 cross-validation. Note. SN = Sensitivity; PPV = Precision; OTH = Other RBC types; BUR = Burr cells; FRA = Fragmented RBCs; NOR = Normal RBCs; OVA = Ovalocytes; TEA = Teardrops; THR = Three overlapped RBCs; TWO = Two overlapped RBCs.

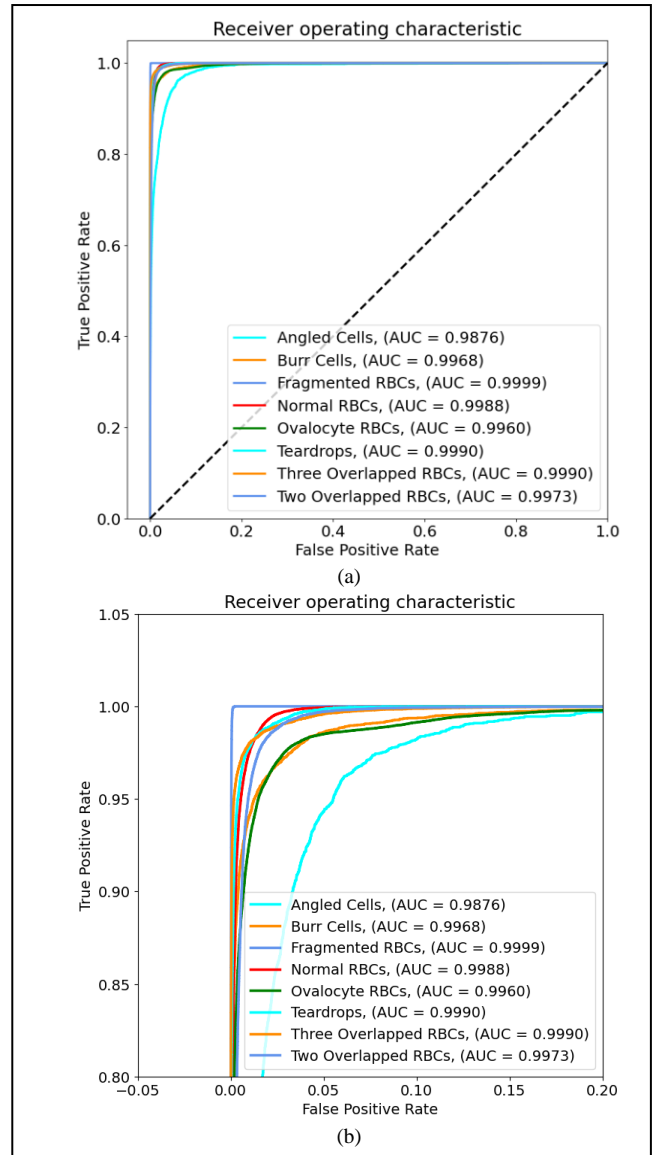

**Fig. 7.** Averaged ROC curves for the proposed RBC classification model: (a) Full-range ROC curves with a range of [0,1] for the true-positive rate and the false-positive rate, (b) Magnified low-range ROC curves with a TPR range of [0.8, 1] and a FPR range of [0, 0.2].

**Table 5**
Performance comparison against eight state-of-the-art CNN models.

| Model | ACC(%) | SN(%) | PPV(%) | F1(%) | Training time (sec.) | Testing time on a GPU (sec.) | Trainable params (M) | Storage size (MB) |
|---|---|---|---|---|---|---|---|---|
| **MobileNetV3Small** | 93.4 | 88.8 | 90.9 | 89.7 | 1,953 | 9.5 | 0.90 | 11.3 |
| **MobileNetV3Large** | 95.5 | 92.0 | 93.3 | 92.4 | 2,072 | 12.5 | 2.90 | 35.0 |
| **ResNet50V2** | 95.7 | 92.0 | 92.7 | 92.3 | 3,406 | 17.7 | 23.50 | 270.2 |
| **ResNet50** | 95.9 | 92.3 | 93.1 | 92.6 | 3,667 | 17.7 | 23.55 | 270.4 |
| **DenseNet121** | 96.3 | 93.6 | 94.1 | 93.8 | 4,563 | 21.2 | 6.96 | 81.5 |
| **Xception** | 96.7 | 94.3 | 94.6 | 94.4 | 4,046 | 18.3 | 20.82 | 239.1 |
| **ConvNeXtTiny** | 96.7 | 94.3 | 94.8 | 94.5 | 5,463 | 40.4 | 27.8 | 319.1 |
| **EfficientnetV2B0** | 96.6 | 94.1 | 94.6 | 94.3 | 3,191 | 13.6 | 5.80 | 68.4 |
| **Proposed method** | 96.5 | 94.1 | 94.3 | 94.2 | 3,378 | 11.0 | 4.02 | 47.0 |

Note. ACC = Test accuracy; SN = Sensitivity; PPV = Positive predictive value or precision; F1 = F1-score.

**Table 6**
Statistical significance test results for the proposed model against eight state-of-the-art CNN models. The p-values with solid underlines indicate significant improvements of our proposed method against the SOTA CNN models. The p-values with dotted underlines indicate significant improvements in the SOTA CNN models against our proposed method. The other p-values indicate insignificant differences.

| Model | ACC | SN | PPV | F1 | Training time (sec.) | Testing time on a GPU (sec.) |
|---|---|---|---|---|---|---|
| **MobileNetV3Small** | 3.15E-13 | 2.62E-13 | 7.29E-11 | 5.30E-13 | 1.18E-15 | 5.88E-03 |
| **MobileNetV3Large** | 1.80E-02 | 1.94E-02 | 7.25E-04 | 1.07E-02 | 3.06E-09 | 3.61E-01 |
| **ResNet50V2** | 8.03E-06 | 1.53E-08 | 9.17E-06 | 3.17E-08 | 6.74E-01 | 5.12E-05 |
| **ResNet50** | 4.67E-04 | 4.47E-06 | 1.83E-04 | 9.70E-06 | 3.36E-02 | 4.57E-05 |
| **DenseNet121** | 1.85E-01 | 1.54E-02 | 2.33E-01 | 2.52E-02 | 8.64E-12 | 9.01E-09 |
| **Xception** | 4.46E-02 | 7.20E-02 | 1.79E-01 | 3.98E-02 | 7.97E-05 | 1.40E-05 |
| **ConvNeXtTiny** | 8.09E-03 | 1.76E-01 | 6.14E-02 | 3.50E-02 | 4.48E-11 | 3.89E-11 |
| **EfficientnetV2B0** | 3.90E-01 | 5.28E-01 | 3.21E-01 | 2.74E-01 | 1.72E-02 | 2.50E-02 |

Note. ACC = Test accuracy; SN = Sensitivity; PPV = Positive predictive value or precision; F1 = F1-score.

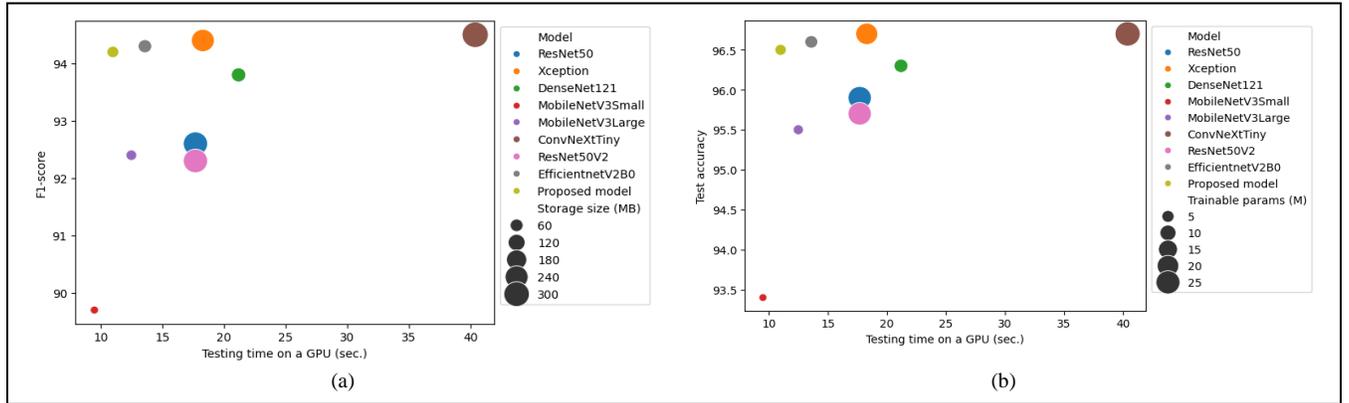

**Fig. 8.** Visual comparison of the classification performance and computational cost measures of the proposed model against eight state-of-the-art CNN models. (a) F1-score, storage size, and testing time on a GPU, and (b) Test accuracy, number of trainable parameters, and testing time on a GPU.

### 5.3. Ablation study

In this section, we conduct an ablation study to investigate the effects of cell segmentation and class-weighted loss on the proposed RBC classification model. First, we created a baseline model with unweighted loss and no cell segmentation. That is, we bypassed the segmentation step of the proposed method and trained a model with raw un-segmented images. Second, we created a variant of the baseline model with cell segmentation to investigate how RBC classification is affected by the presence or absence of neighboring structures. Third, we replaced the unweighted loss by the class-weighted classification loss to address the class imbalance problem. The classification performance results of the three variants were compared using a 5×2 cross-validation scheme. Table 7 shows the average performance evaluation results and statistical significance testing outcomes for these three model variants. The results reveal that the proposed method with segmentation is significantly better than the proposed method without segmentation according to the four performance measures. Segmentation led to a small improvement of almost 2% in the F1-score. This small difference can be ascribed to the good quality of the data where each targeted cell is centered in the middle of an 80×80 image. This ensured model robustness even when fed by un-segmented images. Moreover, the test results show that the proposed method with un-weighted loss is significantly better than the proposed model with class-weighted loss in terms of the test accuracy, precision, and F1-score. The class-weighted loss improves the sensitivity but reduces the precision for the minority classes.

*5.4. Comparison of model performance per slide*

In this section, we show the performance of the proposed model on the test images (17,954 images) for each of the test slides. Table 8 presents the average performance measures of the proposed model for five slides. Even though the model was trained by images from all scanners, the results show variations in performance. The test accuracy, sensitivity, precision, and F1-score are computed for the proposed RBC classification model per slide using a 5×2 cross-validation scheme. Slides 13 and 16 acquired by Scanner 4 show the lowest F1-score because of the blur pattern variations for this scanner. Meanwhile, Slides 5 and 6 acquired by the high-quality Scanner 1 achieve the highest F1-score. Also, images from Slide 4 achieve a good F1-score of 92.05% because of the high quality of Scanner 2.

## 6. Discussion

### 6.1. Summary

In this paper, we introduced a large-scale RBC image dataset for RBC segmentation and classification. This dataset included wide real-world variations of RBC cell images categorized into eight different classes. To the best of our knowledge, this is the largest, most diverse image dataset for RBC classification. It's nearly five times larger than the largest publicly available dataset for RBC classification, namely, the Chula RBC-12 Dataset [24]. Beside the significant difference in the dataset size, our dataset contains images collected from four different scanners, and this adds more data variations with rich details.

Based on the introduced dataset, we built a deep-learning-based two-stage model for RBC segmentation and classification. The RBC segmentation module was based on a U-Net architecture, while the classification module was based on an EfficientNetB0 architecture. We selected a U-Net model for RBC segmentation because of the wide adoption of this model in several applications of medical image segmentation. Also, the simple design of U-Net, paired with skip connections, helps minimize the number of parameters, resulting in less inference time without sacrificing segmentation accuracy. Moreover, the skip connections facilitate fusion of low-level and high-level features, and this allows the model to capture fine-grained details and global context.

**Table 7**
Average performance evaluation results for the proposed RBC classification model variants using a 5×2 cross-validation scheme. Statistical significance test results (based on the two-sample t-test) are shown in brackets for the proposed model vs the proposed method + class-weighted loss, and the proposed model without segmentation.

| Model | Segmentation | Class-weighted loss | ACC (p-value) | SN (p-value) | PPV (p-value) | F1 (p-value) |
|---|---|---|---|---|---|---|
| **proposed method + unweighted loss** | × | × | 95.5% (1.25E-07) | 92.7% (4.46E-06) | 92.4% (2.54E-05) | 92.5% (8.15E-07) |
| **proposed method + segmentation + unweighted loss** | ✓ | × | 96.5% | 94.1% | 94.3% | 94.2% |
| **proposed method + segmentation + class-weighted loss** | ✓ | ✓ | 95.7% (2.87E-04) | 94.6% (2.79E-03) | 91.8% (2.10E-07) | 93.0% (5.77E-05) |

Note. ACC = Test accuracy; SN = Sensitivity; PPV = Positive predictive value or precision; F1 = F1-score.

**Table 8**
Average performance measures (%) of the proposed RBC classification model for the five slides (patients). Each measure is averaged for the 5×2 cross-validation scheme results.

| WSI no. | Scanner type | # of cell images | ACC (%) | SN (%) | PPV (%) | F1 (%) |
|---|---|---|---|---|---|---|
| 4 | Scanner 2 | 3,559 | 96.07 | 92.44 | 91.92 | 92.05 |
| 5 | Scanner 1 | 4,150 | 97.06 | 95.13 | 95.54 | 95.29 |
| 6 | Scanner 1 | 4,678 | 97.32 | 95.63 | 96.22 | 95.89 |
| 13 | Scanner 4 | 3,046 | 95.40 | 82.35 | 83.48 | 82.79 |
| 16 | Scanner 4 | 2,521 | 95.79 | 91.45 | 90.06 | 90.30 |

Note. ACC = Test accuracy; SN = Sensitivity; PPV = Positive predictive value or precision; F1 = F1-score.

The main reason for selecting the EfficientNetB0 model is the challenging nature of the RBC classification problem which requires a model with the ability to achieve high accuracy at a low computational cost. These requirements are strengthened by the fact that each WSI requires extensive analysis of thousands of RBC images in clinical practice.

*6.2. Visual segmentation results*

For RBC segmentation, the trained segmentation model achieves high performance reaching an IoU of 98.03%. Fig. 9 shows samples of the successful segmentation results. From left to right, we show samples of the input images, the ground-truth segmentation maps, the predicted segmentation maps, and the segmented output images. Although these images are challenging, the proposed model could segment the RBC images correctly.

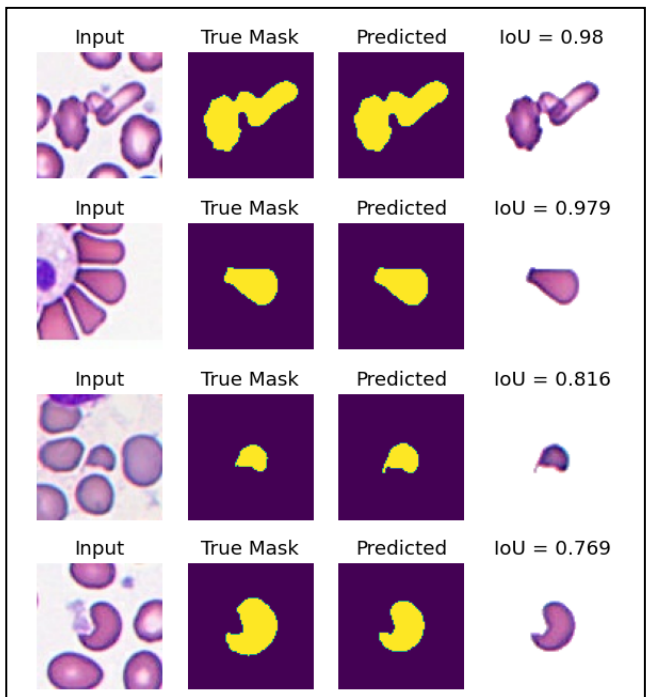

**Fig. 9.** Samples of the RBC segmentation results. The columns from left to right show the original input images, the ground-truth segmentation maps, the segmentation maps predicted using our U-Net-based model, and the segmented output images, respectively.

*6.3. Challenging segmentation cases*

Most of the challenging samples for any segmentation model (including ours) are those with overlapped cells or challenging backgrounds. In Fig. 10, samples with challenging cells are shown with corresponding ground-truth and predicted segmentation maps. For the image sample (a), parts of the cell are covered by a neighboring object similar to the background. This caused our segmentation model to assume the contour of the occluded cell. Still, our segmentation model correctly identified large portions of the cell.

For image sample (b), the target cell comes with an irregular contour shape that touches parts of a neighbor cell. This is also challenging for manual segmentation to identify the actual cell border. However, our model correctly identified a large area of the cell. For image sample (c), small parts of the cell are not visible with color similar to the background. As a result, our model missed these parts and got an IoU of 87.4%.

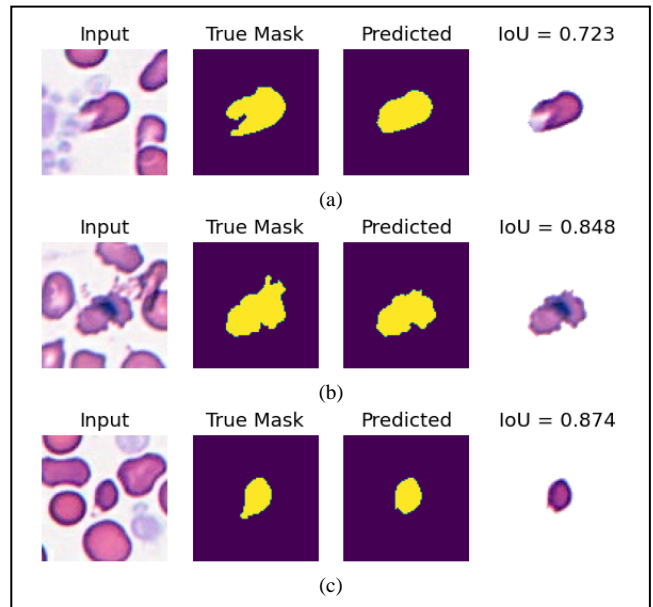

**Fig. 10.** Samples of the RBC segmentation results for challenging cells. The columns from left to right show the original input images, the ground-truth segmentation maps, the segmentation maps predicted using our U-Net-based model, and the segmented output images, respectively.

*6.4. Challenging classification cases*

In the RBC classification experiments, although the sensitivity of the proposed model is high for both normal RBCs and ovalocytes, we notice a high confusion between these types. This confusion can be ascribed to the high similarity between the two types in terms of roundness, overall shape, and texture. Fig. 11(a-c) show correct predictions and Fig. 12(a-c) show misclassified cells.

Also, our model faces challenges in correctly identifying cells of other RBC types (Class 4). Indeed, due to the training data scarcity for this class, our model exhibits the lowest sensitivity (76.1%) for identifying samples of this class (in comparison to other classes). In particular, Fig. 12(d-f) show samples where confusion occurred between this class and each of the ovalocyte and teardrop classes. In Fig. 11(d-f), correct prediction samples for other RBC types are presented.

Moreover, the proposed model faces challenges in classifying samples with two or three overlapped cells. In Fig. 11(g-i), we present correctly predicted samples of two and three overlapped cells. Fig. 12(g-i) show misclassified samples where the classifier confused the two-cell and three-cell overlapped patterns. This confusion can occur due to segmentation errors or the similarity of the contour and texture patterns.

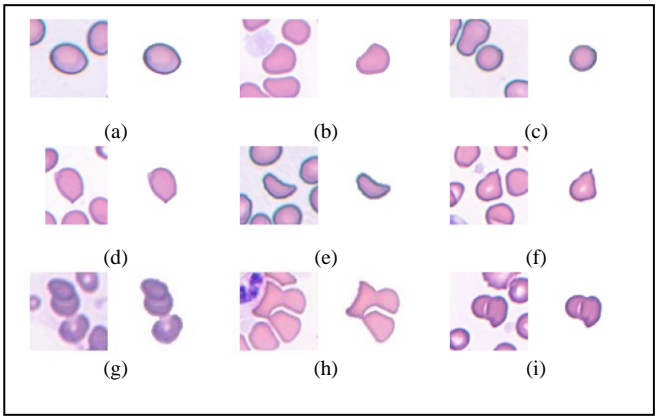

**Fig. 11.** Samples for successful RBC classification results (where each pair consists of a cell image (left) and its corresponding segmentation output (right)). (a, b) Ovalocyte RBCs, (c) Normal RBCs, (d, e, f) Other RBC types, (g, h) Three overlapped RBCs, and (i) Two overlapped RBCs.

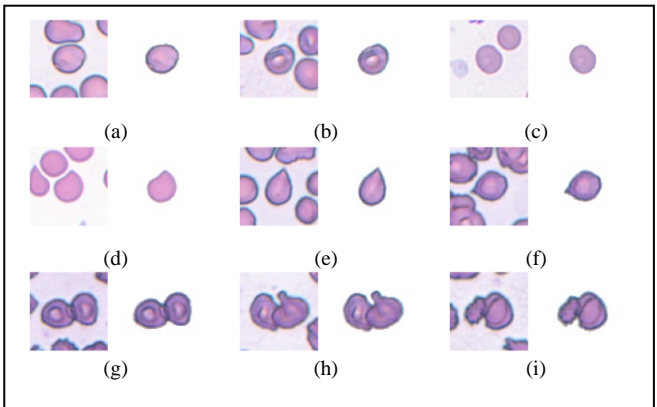

**Fig. 12.** Challenging samples for RBC classification results (where each pair consists of a cell image (left) and its corresponding segmentation output (right)). (a) True label: normal RBCs, predicted: ovalocyte RBCs, (b, c) True label: ovalocyte RBCs, predicted: normal RBCs, (d) True label: ovalocyte RBCs, predicted: other RBC types, (e, f) True label: other RBC types, predicted: teardrops, (g, h) True label: two overlapped RBCs, predicted: three overlapped RBCs, and (i) True label: three overlapped RBCs, predicted: two overlapped RBCs.

### 6.5. Clinically significant RBCs

Abnormal cells identified and counted by the classifier are typically clinically significant RBCs that require visual examination and smear review by hematologists. In addition, there are no current alternative technological or automated analyzers for detection and counting of teardrop RBCs, ovalocytes, or burr cells. Fragmented RBCs and teardrop RBCs are present exclusively in serious medical conditions, where the former is critical to be counted and reported for confirming the diagnosis of relevant diseases. Overlooking these types of RBCs can be fatal, especially prior to platelet transfusion. The teardrop RBCs are present in bone marrow fibrosis and other hematopoiesis disorders including neoplastic disease, bone marrow cancer, and severe anemia under benign conditions. The absence of the teardrop RBCs is a good negative indicator that helps in the exclusion and prognosis of such diseases. Moreover, the percentage of ovalocytes is increased by about 5-10% in all types of anemia.

Teardrops, fragmented RBCs, and ovalocytes are clinically significant and shall not be misclassified by the designed classifiers. The trained classifiers shall indeed be empowered to exclude the artifacts, false teardrops, and false fragmented RBCs by including the burr cells and other RBC types within the training pool as separate classes. These critical pitfalls are mitigated in the proposed classifier by using a dataset that includes the main clinically significant RBC types (fragmented RBCs, teardrops, and ovalocytes). Moreover, without using the class-weighted loss, the proposed model can differentiate between these three types of clinically significant RBCs, and the most confusing classes (namely, the burr cells, and the other RBC types). As well, high sensitivity and precision are obtained for the clinically significant RBCs.

### 6.6. Comparison with state-of-the-art methods

In comparison with other methods utilizing state-of-the-art CNN models, our proposed framework shows a good accuracy-speed balance with a high classification accuracy and low computational cost. Indeed, our method classifies 17,954 test images with high accuracy of 96.5% and the second lowest overall test time of 11 seconds on a GPU. In addition, the proposed model with the class-weighted loss shows the ability to improve the sensitivity for the minority classes with a minor reduction in the overall accuracy to 95.7%. The proposed model is the most appropriate one among the compared RBC classification models because of its ability to reach high classification accuracy and also very low testing time. This shall be valuable in clinical practice when hundreds and thousands of RBC images are required for identification.

## 7. Conclusions and Future Directions

### 7.1. General conclusions

We proposed an RBC image analysis framework with two-stage DL models for cell segmentation and classification using a U-Net architecture and an EfficientNetB0 architecture, respectively. Due to clear limitations in the public RBC datasets, we first built what is currently the largest RBC image dataset for RBC segmentation and classification. The results show the ability of the proposed DL framework to achieve high performance in segmentation with an IoU of 98.03%, a classification test accuracy of 96.5%, and a relatively low testing time of 11 seconds on a GPU for a fold of 17,954 test images. Moreover, our proposed model outperforms eight state-of-the-art CNN models by reaching a good balance between performance and computational cost.

### 7.2. Limitations

This work has some limitations. The dataset was collected from WSI images of only a 40X magnification level. The proposed model may show a drop in performance if tested on images of other magnification levels. Also, we investigated only one DL architecture for cell segmentation. A more detailed study may investigate other alternative architectures with opportunities for better overall performance.

### 7.3. Future directions

There are several possible extensions of this work for improving performance. The introduced dataset comes with wide variations because of the staining and scanning processes. To normalize network inputs, advanced preprocessing and normalization methods (such as stain normalization) should be applied. Also, the proposed DL models can be further refined using better optimized hyperparameters or alternative architectures. Moreover, the dataset can be further enriched with additional images to reflect more diverse real-world conditions and reduce class imbalance.

## Data availability

The 100K-RBC-PathOlOgics and 100K-RBC-Mask-PathOlOgics datasets used in this work represent an early version of the recently published PathOlOgics_RBCs datasets [47].

## Declaration of competing interest

The second and third authors (AE & YA) are cofounders of PathOlOgics, a limited-liability company for processed medical data and tele-pathology services. The other authors have no relevant financial or non-financial interests to disclose.

## Funding

The authors declare that no funds, grants, or other support were received during the preparation of this manuscript.